\documentstyle[twocolumn,twoside,floats,prb,aps,psfig]{revtex}

\begin{document}
\newcommand{\beq}{\begin{equation}}
\newcommand{\eeq}{\end{equation}}

\title{A Rational Approach to Ring Flexibility in Internal Coordinate
Dynamics}
\author{Alexey K. Mazur}
\address{Laboratoire de Biochimie Th\'eorique, CNRS UPR9080\\
Institue de Biologie Physico-Chimique\\
13, rue Pierre et Marie Curie, Paris,75005, France.\\
FAX:(33-1) 43.29.56.45. Email: alexey@ibpc.fr}

\address{\medskip\em
\begin{minipage}{14cm}
{}~~~
Internal coordinate molecular dynamics (ICMD) is an efficient method
for studying biopolymers, but it is readily applicable only to
molecules with tree topologies, that is with no internal flexible
rings. Common examples violating this condition are prolines and
loops closed by S-S
bridges in proteins. The most important such case, however, is
nucleic acids because the flexibility of the furanose rings always plays
an essential role in conformational transitions both in DNA and RNA.
There are a few long-known theoretical approaches to this problem, but,
in practice, rings with fixed bond lengths are closed by adding
appropriate harmonic distance restraints, which is not always
acceptable especially in dynamics. This paper tries to overcome this
handicap of ICMD and proposes a rational strategy which results in
practical numerical algorithms. It gives a unified analytical
treatment which shows that this problem is very close to the
difficulties encountered by the method of constraints in Cartesian
coordinate dynamics, and certain ideas of the latter appear helpful
in the context of ICMD. The method is affordable for large molecules
and generally applicable to all kinds of rings. A specific
implementation for five-membered rings is described and tested for a
proline-rich polypeptide and a decamer DNA duplex. In both cases
conditions are found which make possible time steps around 10 fsec in
ICMD calculations.
\end{minipage} }

\maketitle
\section{Introduction}
Internal coordinate molecular dynamics
\cite{BKS0:89,Jain:93,Rice:94,Mzjcc:97}
(ICMD, see Ref. \onlinecite{Mzjcc:97} for a historical review) is a
recent approach
in the simulation of flexible polymers which, unlike the traditional one,
employs torsions and, if desired, valence angles and bond lengths as
generalized coordinates in the equations of motion. It originates from the
Euler-Lagrange-Hamilton formalism of classical mechanics and makes
possible modeling of polymers as chains of rigid bodies, which
automatically eliminates the most severe time step limitations
characteristic for Newtonian MD. In addition, it drastically
reduces the configurational space of flexible molecules, which is very
useful for conformational searches \cite{Mzjbsd:93,Mzcr:95}, and for
refinement of experimental structures by simulated annealing
\cite{Rice:94,Stein:97}.

Treatment of flexible cycles is an inherently difficult task for ICMD.
The possibility of applying this method to large polymers rests upon
recursive algorithms which all are applicable only when the molecule
is topologically isomorphous to a tree
\cite{Vereshchagin:74,Featherstone:87,Rodriguez:87,Rodriguez:92,Jain:91}.
Any non-rigid cycle,
therefore, makes the whole system unsuitable for ICMD. Although a few
possible approaches to this problem can be readily
sketched,\cite{BKS1:89,Rodriguez:91} until now,
no practical solution has been reported. This difficulty is most
critical for simulations of nucleic acids because the five-membered
sugar rings connect the bases with the sugar-phosphate backbone, and
their relative orientations are, therefore, determined by the
pseudorotation states of the sugars. As a result, nucleic acids are
completely beyond the scope of ICMD although one cannot say that the
problem did not attract attention. \cite{Rudnicki:94}

This paper describes a rational solution of the problem of ring
flexibility in ICMD. The new approach has much in common with the
recent methods of constraints in Newtonian dynamics. \cite{LINKS:97}
The similarity between the difficulties created by ring
flexibility in ICMD, and bond length constraints in Newtonian MD,
is intuitively clear. Both these problems can be consistently treated
by using projection operators in linear spaces, which results in a
unified formalism. A fruitful idea used, sometimes implicitly, in
Newtonian constraint dynamics is that
projection operators can be applied directly in the finite-difference
form of the equations of motion. This simplifies computations because
certain terms in the analytical equations can be omitted, and
immediately results in a time-reversible symplectic numerical
integrator.

A specific implementation for the most practically important case of
five-membered rings is described and tested here. It employs our
earlier analytical approach to ring flexibility,
\cite{BKS1:89,Mzc&c:90} but can also be adapted to alternative
methods, for instance, with various pseudorotation equations.
Numerical algorithms suitable for other specific cases, such as S-S
bridges in proteins, are also tested, but such cases are not studied
here in detail. For five-membered rings in proteins and nucleic acids
specific conditions that make possible time steps around 10 fsec are
considered.

\section{Results and Discussion}
\subsection*{Overview of Earlier Approaches}
\begin{figure}
\centerline{\psfig{figure=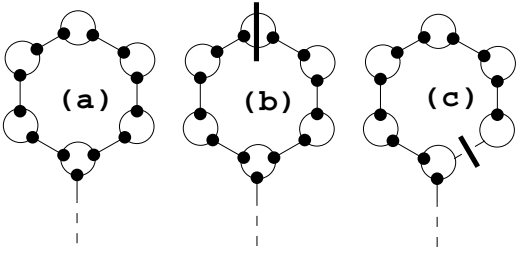,width=7.5cm,angle=0.}}
\caption{\label{Floop}
Schematic representation of a loop of rigid bodies. To impose
a tree topology loop {\em (a)} must be considered as broken, which can be done
in many ways, for example, as in loops {\em (b)} or {\em (c)}. Rigid bodies
shown as open circles are joined by hard sticks in pivots shown as closed
circles.
}\end{figure}

In this section we outline the general view of the problem of ring
flexibility and try to divide it into distinct separable tasks. This
problem has been first addressed by G\= o and Scheraga \cite{Go:70} who
formulated the following general approach later followed by others.
Consider the example in Fig.\ref{Floop}. Suppose we
want to describe a flexible loop {\em (a)} with certain internal
coordinates that
determine variable angles at pivots and lengths of bonds.
A correct definition of internal coordinates always requires that the
molecule is represented as a regular tree, which means that,
conceptually, loop {\em (a)} in Fig.\ref{Floop} must be considered broken
in agreement with a certain tree topology, while a correct loop
closure is maintained by applying appropriate constraining
conditions. Hereafter this construction is referred to as the underlying
tree. The constrained underlying tree gives the desired model with
correctly closed loops, but, in derivations, its unconstrained
movements are also sometimes considered.
Structures {\em (b)} and {\em (c)} in Fig.\ref{Floop}
demonstrate that there are numerous ways to define the underlying tree
of any given loop.

Now suppose we need to sample from the ensemble of closed loop
conformations. Internal coordinates can be denoted as an n-vector $\bf
q$ and they should vary concertedly so that the $k$ scalar
constraining conditions
\begin{equation}\label{Ecnco}
C_\mu({\bf q})=0,\ \mu=1,...,k.
\end{equation}
are not violated.
By inverting Eqs. (\ref{Ecnco}), a set of $k$ internal coordinates can be,
in principle, computed from the rest, which may be written as
\begin{equation}\label{Edpva}
{\bf q^d}={\bf q^d}({\bf q^f})
\end{equation}
where $\bf q^d$ and $\bf q^f$ are the k-vector of
dependent and (n-k)-vector of independent variables, respectively. Now
$\bf q^f$ can be freely varied within the range of solubility of
Eqs. (\ref{Ecnco}). Dependant coordinates $\bf q^d$ computed by Eq.
(\ref{Edpva}) always provide a correct loop closure in the underlying
tree. Time derivatives of Eq. (\ref{Edpva}) read
{\mathletters\label{Edfdp=}\begin{eqnarray}
{\bf\dot q^d}&=&\frac{\partial{\bf q^d}}{\partial{\bf q^f}}{\bf\dot q^f}\\
{\bf\ddot q^d}&=&\frac{\partial{\bf q^d}}
{\partial{\bf q^f}}{\bf\ddot q^f}+{\bf\dot q^f}\frac{d}{dt}\frac{\partial{\bf q^d}}{\partial{\bf q^f}}
\end{eqnarray}}
where $\partial{\bf q^d}/\partial{\bf q^f}$ is a $k\times (n-k)$ matrix.
This approach may be called consistent because, following to the basic
idea of internal coordinates, it reduces the number of independent
variables to the true number of conformational degrees of freedom.

The above reasoning reveals the first task to be considered, that is,
inversion of Eqs. (\ref{Ecnco}) or, in other words, construction of a
correctly closed ring geometry for any given $\bf q^f$.
This is generally difficult because Eqs.
(\ref{Ecnco}) are non-linear. Practical solutions exist only in a few
special cases, notably, there are relatively simple algorithms for
loop closure with flexible valence angles.
\cite{BKS1:89,Mzc&c:90,Lugovskoy:72}
For the important case of five-membered rings, pseudorotation equations
give the most economical solution. \cite{Altona:72} The second task
is the calculation of dependent velocities and accelerations, but
usually this presents no serious difficulties.

The possibility to apply this general strategy in ICMD was first
addressed in Ref. \onlinecite{BKS1:89}. Explicit equations of motion in
independent variables can be obtained by linear transformations
of the familiar ICMD equations for the unconstrained underlying tree.
It is useful to reproduce this result here in a compact form.
Let us construct $n\times (n-k)$ matrix $\bf P$ as \[{\bf
P}=\frac{\partial{\bf q}}{\partial{\bf q^f}}.\] Every column in
$\bf P$ is composed of partial derivatives of all internal
coordinates with respect to a certain independent variable.
Thus, each column has one unit element, $n-k-1$ zeroes and $k$
derivatives of dependent variables. Any
vector $\bf\dot q^f$ yields velocities of the underlying tree
\[{\bf\dot q}={\bf P\dot q^f}\] that fulfill the loop closure
conditions. Similarly, the constrained accelerations are given by
\begin{equation}\label{EGv}
{\bf\ddot q}={\bf P\ddot q^f}+{\bf\dot P\dot q^f}.
\end{equation}

Consider the unconstrained motion of the underlying tree.
The corresponding equations of motion can be written as\cite{BKS0:89}
\begin{equation}\label{EmBKS0}
{\bf M}({\bf q}){\bf\ddot q}={\bf f}({\bf
q})+{\bf u}({\bf q},{\bf\dot q}),
\end{equation}
where ${\bf M}({\bf q})$, ${\bf f}({\bf q})$ and
${\bf u}({\bf q},{\bf\dot q})$ are the mass matrix, the vector of
generalized forces and the inertial term, respectively.
In Ref. \onlinecite{BKS1:89} equations for independent variables
in a correctly constrained underlying tree
were obtained by summing up scalar lines in Eq. (\ref{EmBKS0})
corresponding to dependent variables with coefficients from
matrix ${\bf P}$, and excluding ${\bf\ddot q^d}$ with the help of
Eq. (\ref{Edfdp=}b).
These calculations may be equivalently expressed in a matrix form as
\begin{equation}\label{EmBKS1}
{\bf P}^*{\bf MP\ddot q^f}={\bf
P}^*\left({\bf f}+{\bf u}-{\bf M\dot{P}\dot q^f}\right)
\end{equation}
where the asterisk denotes transposition. All accelerations for the
constrained underlying tree are computed by substituting 
${\bf\ddot q^f}$ into Eq.(\ref{EGv}).
For the following discussion it is convenient to rewrite it as
\begin{eqnarray}\label{Eanrz0}
{\bf\ddot q}=\left[{\bf P}\left({\bf
P}^*{\bf MP}\right)^{-1}{\bf P}^*{\bf M}\right]{\bf M}^{-1}
\left({\bf f}+{\bf u}\right)\nonumber\\
+\left({\bf I}-{\bf P}\left({\bf
P}^*{\bf MP}\right)^{-1}{\bf P}^*{\bf M}\right){\bf\dot P\dot q^f}.
\end{eqnarray}

Equations (\ref{EmBKS1}) and (\ref{Eanrz0}) clearly show that the
performance of any numerical algorithm originating from the consistent
approach is limited by the necessity of inverting the $(n-k)\times (n-k)$
mass matrix ${\bf P}^*{\bf MP}$. This no longer corresponds to
a tree topology therefore the fast recursive algorithms
\cite{Rodriguez:87,Rodriguez:92}
are inapplicable and, at least at present, 
only straightforward inversion is possible, which needs $O\left[(n-k)^3\right]$
computations. In practice a reasonable maximum number of variables
for calculations with a straightforward mass matrix inversion is
around 100. This approach, therefore, is hardly practical for S-S
bridges and prolines in globular proteins. A minimalist
representation of nucleic acids needs about eight degrees of freedom
per nucleotide, which comprises two pseudorotation variables for a
furanose ring \cite{Altona:72,Gabb:95}, one torsion for the base and
all backbone torsions. Thus, a decamer duplex would be quite manageable since
matrices of separate strands are inverted separately, but a hairpin of
the same lengths would be more difficult.

This difficulty could be avoided, in principle, if we did not
divide internal coordinates into dependent and independent variables.
For instance, we could compute all reaction forces in a closed loop
and include them explicitly into equations of motion of the
unconstrained tree, which should give correct closed loop dynamics.
This idea has been in fact elaborated in robot mechanics
\cite{Rodriguez:91} and, at least theoretically, it should result in an
$O(n)$ algorithm. Note also that this would be equivalent
to the introduction of holonomic constraints explicitly, similarly to the
well-known method of constraints in Cartesian coordinate dynamics
\cite{SHAKE:77}. In a certain sense such an approach may be called
inconsistent because the resultant system is constrained both
implicitly and explicitly, but we will see below that this appears
more efficient.

\subsection*{Projection Operators}
In this section we briefly recall a few necessary facts concerning
projection operators in linear spaces. Consider a mechanical system
of $N$ free particles. Its configuration is determined by an
$n$-vector of Cartesian coordinates $\bf x$, where $n=3N$. Forces
applied to particles are given by an $n$-vector $\bf f$. The
possible values of $\bf f$ cover a full $n$-dimensional vector
space ${\cal L}^n$.

Now let us assume that our particles are bound to move along some
hypersurface defined by a constraint equation $$C({\bf x})=0$$
and that initially they are placed at point ${\bf x}_0$
with zero velocities. Vector
$${\bf g}=\nabla C\begin{array}{|l}\\{\bf x}={\bf x}_0\end{array},$$
where $\nabla$ is a multidimensional gradient operator, is orthogonal
to the constraint hypersurface and all its orthogonal vectors form an
$(n-1)$-dimensional subspace in ${\cal L}^n$ called tangent hyperplane
${\cal P}^{n-1}$. The particles are held on the constraint
hypersurface by reaction forces given by vector ${\bf f}^\bot$
collinear to $\bf g$. Together with $\bf f$ this vector must give
${\bf f}^\|\in{\cal P}^{n-1}$: \begin{equation}\label{Ealpha} {\bf
f}^\|={\bf f}+{\bf f}^\bot={\bf f}+\alpha({\bf f})\bf g \end{equation}
where $\alpha({\bf f})$ is a scalar function.

It is clear that Eq. (\ref{Ealpha}) applies to any vector ${\bf
f}\in{\cal L}^n$. It is known also that the sum of any two free
forces ${\bf f}_1+{\bf f}_2$ gives the corresponding sum ${\bf
f}_1^\|+{\bf f}_2^\|$, and that $\bf f$ multiplied by a constant
results in the same multiplication of ${\bf f}^\|$. In other words, Eq.
(\ref{Ealpha}) in fact defines a linear mapping ${\bf
f}\longrightarrow{\bf f}^\|$ from ${\cal L}^n$ to ${\cal P}^{n-1}$,
and, consequently, it can be expressed as
\begin{equation}\label{EdefT}
{\bf f}^\|={\bf Tf}
\end{equation}
where $\bf T$ is an $n\times n$ matrix.

By construction, ${\bf T}^2=\bf T$, that is $\bf T$ represents a
projection operator. Any such matrix can be calculated as
\begin{equation}\label{EclT}
{\bf T}=({\bf I}-{\bf\tilde T})=({\bf I}-{\bf
D}({\bf G}^*{\bf D})^{-1}{\bf G}^*)
\end{equation}
where $\bf I$ is the unit matrix, $\bf D$ and $\bf G$ are basis matrices of
the right and left zero spaces of $\bf T$, respectively. By
definition, vector $\bf e$ belongs to the right zero space of $\bf T$
if ${\bf Te}=0$. Such vectors form a linear subspace ${\cal D}^k$ in
${\cal L}^n$ and $n\times k$ matrix $\bf D$ in Eq. (\ref{EclT})
consists of $k$ its basis vectors. It is readily verified that ${\bf
TD}=0$. Matrix $\bf G$ is constructed in the same way, and it is seen
that ${\bf G}^*{\bf T}=0$. The two zero spaces ${\cal D}^k$ and ${\cal
G}^k$ always have the same dimension and they define a unique
projection operator.

The physical meaning of ${\cal D}^k$ and ${\cal G}^k$ is
clear from the above example. ${\cal D}^k$ determines the direction of
projection. Equation (\ref{Ealpha}) shows that, in our example, the
corresponding basis consists of a single vector $\bf g$. In turn,
${\cal G}^k$ is the orthogonal compliment to the tangent hyperplane
${\cal P}^{n-k}$. In our example ${\cal G}^1$ it is identical to
${\cal D}^1$, and so we have $${\bf T}=1-{\bf g}({\bf g}^*{\bf
g})^{-1}{\bf g}^*$$

Finally, consider
\begin{equation}\label{EclT'}
{\bf\tilde T}={\bf D}({\bf
G}^*{\bf D})^{-1}{\bf G}^*.
\end{equation} It is readily verified
that $\bf\tilde T$ is also a projecting operator. It projects upon
${\cal D}^k$ along the orthogonal subspace of ${\cal G}^k$, that is,
along the tangent hyperplane ${\cal P}^{n-k}$, thus making a
complimentary pair with $\bf T$.
Equations (\ref{EclT}) and (\ref{EclT'}) can also be applied to
a single projection, which gives two alternative representations.
In both cases, the operator is defined by the direction and the
hyperplane of the projection.
However, in case of Eq. (\ref{EclT}) one has to substitute the
direction itself and the orthogonal compliment for the plane,
whereas in Eq. (\ref{EclT'}) the opposite combination is required.

\subsection*{The Analytical Equations of Constraint Dynamics}
In this section we derive the basic equations of motion for constraint
dynamics directly from the above projection considerations. These
equations are well-known, but our main interest is in the reasoning
rather than in the result because similar arguments will later lead us
to the required numerical algorithm.

We consider again a system with $n$ degrees of freedom
described by an $n$-vector $\bf q$ and the equations of motion (\ref{EmBKS0}).
We assume that these equations can be somehow
derived and solved without major difficulties. For Newtonian MD these
are equations for unconstrained molecules while in the case of ICMD
these are equations for trees of rigid bodies. These two models thus play
similar roles in our reasonings.

Now let us impose on our system a certain number of explicit
constraints defined by Eq. (\ref{Ecnco}). For clarity, from now on we
drop superscripts denoting dimensions of subspaces. Equations
(\ref{Ecnco}) define an
(n-k)-dimensional hypersurface in $\cal L$ and the system is held
on it by generalized reactions ${\bf f}^\bot\in \cal G$ where
$\cal G$ is the subspace of orthogonal vectors $${\bf g}_\mu=\nabla
C_\mu({\bf q})$$ with an $n\times k$ basis matrix $\bf G$. A
straightforward derivation of equations of motion for the constrained
system would require evaluation of ${\bf f}^\bot$ with subsequent
substitution into the r.h.s. of Eq. (\ref{EmBKS0}):
\begin{equation}\label{EcBKS}
{\bf M\ddot q}={\bf f}+{\bf u}+{\bf f}^\bot.
\end{equation}
Calculation of reactions can be bypassed as follows.

By taking time derivatives of Eq. (\ref{Ecnco}) one obtains
constraining conditions upon the tree velocities and
accelerations
{\mathletters\label{Edfcn}\begin{eqnarray}
{\bf g}_\mu{\bf\dot q}&=&0\\
{\bf g}_\mu{\bf\ddot q}+{\bf\dot g}_\mu{\bf\dot q}&=&0
\end{eqnarray}}
Let us first consider the case ${\bf\dot g}_\mu=0$. One can assume, for
instance, that particles are moving along space-fixed surfaces. In
this case the second term in Eq. (\ref{Edfcn}b) is zero and we see
that the constrained accelerations belong to the tangent hyperplane
$\cal P$. Note also that according to Eq. (\ref{EcBKS}) constrained
accelerations are obtained by correcting the corresponding
unconstrained vector by ${\bf M}^{-1}{\bf f}^\bot$, that is, by a
vector from a subspace with the basis matrix ${\bf M}^{- 1}\bf G$. We
see, therefore, that calculation of the constrained accelerations
is nothing but a projection ${\cal L}\longrightarrow{\cal P}$
along direction ${\bf M}^{- 1}\bf G$. The corresponding operator
is readily computed according to Eq. (\ref{EclT})
\beq\label{ET=}
{\bf T}=({\bf I}-{\bf\tilde T})=\left[{\bf I}-{\bf M}^{-1}{\bf G}
\left({\bf G}^*{\bf M}^{-1}{\bf G}\right)^{-1}{\bf G}^*\right]
\eeq
By applying it in Eq. (\ref{EcBKS}) we obtain equations of motion
\begin{eqnarray}\label{Eanrz1}
{\bf\ddot q}&=&{\bf TM}^{-1}({\bf f}+{\bf u})=\\
&&\left[{\bf I}-{\bf
M}^{-1}{\bf G}\left({\bf G}^*{\bf M}^{-1}{\bf G}\right)^{-1}{\bf
G}^*\right]{\bf M}^{-1}({\bf f}+{\bf u}).\nonumber\end{eqnarray}

Now consider the case ${\bf\dot g}_\mu\not= 0$. Equation (\ref{Edfcn}b)
indicates that, unlike velocities, the constrained accelerations no
longer belong to $\cal P$, but lie in another hyperplane which
is shifted from the zero of the coordinates. It does not, therefore,
represent a subspace in $\cal L$ and the constrained accelerations
can no longer be obtained by a linear mapping like Eq.(\ref{EdefT}).
Let us, however, decompose vector $\bf\ddot q$ as
\begin{equation}\label{Eq+q}
{\bf\ddot q}={\bf\ddot q}^\|+{\bf\ddot q}^\bot
\end{equation}
where ${\bf\ddot q}^\|\in{\cal
P}$ and ${\bf\ddot q}^\bot\in\cal G$. By substituting Eq.
 (\ref{Eq+q}) into Eq.(\ref{EcBKS}) we get
\begin{equation}\label{EcBKS-Mq}
{\bf M\ddot q}^\|={\bf f}+{\bf u}-{\bf M\ddot q}^\bot+{\bf f}^\bot.
\end{equation}
Now we can again use operator ${\bf T}$ to compute ${\bf\ddot q}^\|$
and, by substituting back to Eq. (\ref{Eq+q}), obtain
\beq\label{ETM+Tq}
{\bf\ddot q}={\bf TM}^{-1}\left({\bf
f}+{\bf u}\right)+{\bf\tilde T\ddot q}^\bot
\eeq
Thus, we still avoid
explicit calculation of reactions if vector ${\bf\ddot q}^\bot$ is
obtained separately. This is, fortunately, the case. By definition,
${\bf\ddot q}^\bot$ is a projection of $\bf\ddot q$ upon $\cal
G$ along hyperplane $\cal P$. The corresponding
operator can be computed by Eq. (\ref{EclT'}). (Note that
it differs from ${\bf\tilde T}$ in Eq. (\ref{ET=})
by its target hyperplane.) By using Eqs. (\ref{EclT'})
and (\ref{Edfcn}b) we get
\begin{equation}\label{ET'q}
{\bf\ddot
q}^\bot={\bf G}\left({\bf G}^*{\bf G}\right)^{-1}{\bf G}^*{\bf\ddot
q}=-{\bf G}\left({\bf G}^*{\bf G}\right)^{-1}{\bf\dot G}^*{\bf\dot q}.
\end{equation}
and substitution of Eqs. (\ref{ET=}) and (\ref{ET'q}) into
Eq. (\ref{ETM+Tq}) gives the required equation of motion
\beq\label{Eanrz2}\begin{array}{l}
{\bf\ddot q}=\\
\left[{\bf I}-{\bf M}^{-1}{\bf G}\left({\bf G}^*{\bf M}^{-1}{\bf
G}\right)^{-1}{\bf G}^*\right]{\bf M}^{-1}({\bf f}+{\bf u})-\\
{\bf M}^{-1}{\bf G}\left({\bf G}^*{\bf M}^{-1}{\bf G}\right)^{-1}{\bf\dot
G}^*{\bf\dot q}
\end{array}\eeq
Note that for Cartesian coordinates
this gives the same equation as that derived by using Lagrange
multipliers. \cite{LINKS:97} On the other hand, it is seen from Eqs.
(\ref{EGv}) and (\ref{Eq+q}) that $${\bf\ddot
q}^\bot={\bf\dot{P}\dot q^f}$$ and thus  Eq. (\ref{Eanrz0})
is equivalent to Eqs. (\ref{ETM+Tq}) and (\ref{Eanrz2}).
They involve the same projection
operation, but employ two alternative representations of the operator.
These two representations correspond to two alternative approaches to
constraints in dynamics and dictate opposite trends in computational
strategies. In case of Eq. (\ref{Eanrz0}) the projection operator is
specified by the target plane and the orthogonal compliment to the
projecting direction, that is according to Eq. (\ref{EclT'}). This
leads to inversion of a $(n-k)\times (n-k)$ matrix and prompts
a reduction of the degrees of freedom in the system. In contrast, in the case
of Eq. (\ref{Eanrz2}) the operator is specified by the direction and
the orthogonal compliment to the target plane, that is corresponding
to Eq. (\ref{EclT}), which results in inversion of a $k\times k$ matrix
and prompts reduction of the number of constraints and, consequently,
keeping a possibly large number of degrees of freedom. In the second
case, matrix ${\bf M}^{-1}$ is used several
times, but this is not computationally limiting and in practice
Eq. (\ref{Eanrz2}) appears applicable to large systems.

Let us finally obtain quasi-Hamiltonian equations for the same system.
For the unconstrained underlying tree these equations can be expressed as
\cite{Mzjcc:97}
{\mathletters\label{EqH}\begin{eqnarray}
{\bf\dot p}&=&
{\bf f}({\bf q})+{\bf w}({\bf q},{\bf\dot q})\\
{\bf\dot q}&=&{\bf M}^{-1}{\bf p}
\end{eqnarray}}
where $\bf p$ is the n-vector of conjugate momenta
and ${\bf w}({\bf q},{\bf\dot q})$ is the corresponding inertial
term. In order to get the equations for the constrained
tree we just need to evaluate generalized reactions from
Eqs. (\ref{EcBKS}) and (\ref{Eanrz2}) and add them to the r.h.s. of Eq.
(\ref{EqH}a). The vector of reaction is
\begin{equation}\label{Efr=}
{\bf f}^\bot=-{\bf G}\left({\bf G}^*{\bf M}^{-1}{\bf
G}\right)^{-1}\left[{\bf\dot G}^*{\bf\dot q}+{\bf G}^*{\bf M}^{-1}({\bf
f}+{\bf u})\right],
\end{equation}
and the resultant equations read
{\mathletters\label{EqHc}\begin{eqnarray}
{\bf\dot p}&=&\left[{\bf I}-{\bf G}\left({\bf G}^*{\bf M}^{-1}{\bf
G}\right)^{-1}{\bf G}^*{\bf M}^{-1}\right]{\bf f}+{\bf w}-\nonumber\\&&{\bf
G}\left({\bf G}^*{\bf M}^{-1}{\bf G}\right)^{-1}\left({\bf\dot G}^*{\bf\dot
q}+{\bf G}^*{\bf M}^{-1}{\bf u}\right)\\
{\bf\dot q}&=&{\bf M}^{-1}{\bf p}.
\end{eqnarray}}
The last equations are
preferred because they are suitable for symplectic numerical
integration and provide much better average properties of dynamic
trajectories with large time steps.

\subsection*{Numerical Algorithm}
Equations (\ref{EqHc}) have the same general form as Eq. (\ref{EqH})
and are suitable for the implicit leapfrog integrator
designed for the latter.\cite{Mzjcc:97} However, here I prefer to
follow the well-tested approach of the Newtonian MD, which
avoids a straightforward integration of the analytical equations,
and uses projection considerations to derive a high
precision numerical algorithm.\cite{LINKS:97,SHAKE:77} The main idea
becomes clear from the
following example. Consider an Euler step in Cartesian coordinates:
{\mathletters\label{Ecclf}\begin{eqnarray}
{\bf v}_1 &=& {\bf
v}_0+{\bf M}^{-1}\left({\bf f}_0+{\bf f}_0^\bot\right)h\\
{\bf x}_1&=& {\bf x}_0+{\bf v}_1h
\end{eqnarray}}
where $h$ is the step size and subscripts refer to the time step number.
Suppose coordinates ${\bf x}_0$ and velocities ${\bf
v}_0$ satisfy constraints with absolute accuracy.
Equation (\ref{Ecclf}a) with an exact reaction ${\bf f}^\bot_0$
computed by using Eq. (\ref{Efr=}),
for instance, would give velocities ${\bf v}_1$ with an approximation
error of $O(h^2)$. A similar error would be propagated to the constraint
conditions at the next step. Note, however, that, according to Eq.
(\ref{Ecclf}b), we can minimize
deviation of the trajectory from the next step constraint hyperplane
${\cal P}_1$ by requiring that ${\bf v}_1\in{\cal P}_0$.
Effectively, this means that, instead of
the true reaction force ${\bf f}^\bot_0$, we substitute into Eq.
(\ref{Ecclf}a) another vector which provides a projection of ${\bf
v}_0+{\bf M}^{-1}{\bf f}_0h$ upon ${\cal P}_0$.
Algorithm (\ref{Ecclf}), therefore, becomes
{\mathletters\label{EcclfT}\begin{eqnarray}
{\bf v}_1&=&{\bf T}_0\left({\bf v}_0+{\bf M}^{-1}{\bf f}_0h\right)\\
{\bf x}_1&=&{\bf x}_0+{\bf v}_1h
\end{eqnarray}}
where projection operator ${\bf T}_0$ is computed from ${\bf
x}_0$. This basic idea is used in the SHAKE\cite{SHAKE:77}
and LINKS\cite{LINKS:97} algorithms of
Cartesian MD where unconstrained bond lengths are corrected by using
previous bond directions.
The projection in Eq. (\ref{EcclfT}a) does not eliminate the
error in the constraints, but only gives the best first approximation.
The residual error can be controlled by feedback schemes which
add a small out-of-plane correction to the projected vector ${\bf
v}_1$ so as to eliminate accumulation.\cite{LINKS:97}

Now consider the implementation of the above strategy for Eqs. (\ref{EqHc}).
With our present notation the kernel of the implicit leapfrog integrator
for Eq. (\ref{EqH}) reads\cite{Mzjcc:97}
{\mathletters\label{Elf}
\begin{eqnarray}
{\bf f}_n&=&{\bf f}({\bf q}_n)\\
\label{Elfb}\circ\ {\bf q}_{n+\frac 12}&=&{\bf q}_{n-\frac 12}+
\left({\bf\dot q}_{n-\frac 12}+{\bf\dot q}_{n+\frac 12}\right)\frac h2\\
\label{Elfc}\circ\ {\bf p}_{n+\frac 12}&=&{\bf p}_{n-\frac 12}+{\bf f}_nh+
\left({\bf w}_{n-\frac 12}+{\bf w}_{n+\frac 12}\right)\frac h2\\
\circ\ {\bf\dot q}_{n+\frac 12}&=&{\bf M}^{-1}_{n+\frac 12}
{\bf p}_{n+\frac 12}\\
{\bf q}_{n+1}&=&{\bf q}_n+{\bf\dot q}_{n+\frac 12}h
\end{eqnarray}}
where the conventional notation is used for denoting on-step and half-step
values. The lines marked by circles are iterated until convergence of
Eqs. (\ref{Elfb}) and (\ref{Elfc}). Reactions depend upon both
coordinates and velocities, therefore, if computed explicitly, they
should have been added to Eq. (\ref{Elfc}) to give
\begin{eqnarray*}
{\bf p}_{n+\frac 12}&=&{\bf p}_{n-\frac 12}+{\bf
f}_nh+\left({\bf w}_{n-\frac 12}+{\bf w}_{n+\frac 12}\right)\frac
h2+\\
&&\left({\bf f}^\bot_{n-\frac 12}+{\bf f}^\bot_{n+\frac 12}\right)\frac h2
\end{eqnarray*}
Following the above strategy we require that ${\bf f}^\bot_{n+\frac
12}$ provide a projection of the unconstrained predicted velocities upon
${\cal P}_{n+\frac 12}$, which results in the following sequence of
calculations
{\mathletters\label{Elfcn}
\begin{eqnarray}
{\bf f}_n&=&{\bf f}({\bf q}_n)\\
\circ\ {\bf q}_{n+\frac 12}&=&{\bf q}_{n-\frac 12}+
\left({\bf\dot q}_{n-\frac 12}+{\bf\dot q}_{n+\frac 12}\right)\frac h2\\
\circ\ {\bf\tilde p}_{n+\frac 12}&=&{\bf p}_{n-\frac 12}+{\bf f}_nh+\nonumber\\
&&\left({\bf w}_{n-\frac 12}+{\bf w}_{n+\frac 12}\right)\frac h2+
{\bf f}^\bot_{n-\frac 12}\frac h2\\
\circ\ {\bf\dot q}_{n+\frac 12}&=&{\bf T}_{n+\frac 12}{\bf M}^{-1}_{n+\frac 12}
{\bf\tilde p}_{n+\frac 12}\\
{\bf p}_{n+\frac 12}&=&{\bf M}_{n+\frac 12}{\bf\dot q}_{n+\frac 12}\\
{\bf f}^\bot_{n-\frac 12}\frac h2&=&{\bf p}_{n+\frac 12}-{\bf\tilde
p}_{n+\frac 12}\\
{\bf q}_{n+1}&=&{\bf q}_n+{\bf\dot q}_{n+\frac 12}h
\end{eqnarray}}
Equations (\ref{Elfcn}) represent the kernel of the
new algorithm. The prediction step necessary to enter the iterative
cycle of Eqs. (\ref{Elfcn}b-\ref{Elfcn}d) is made by using the previous
half-step values instead of ${\bf\dot q}_{n+\frac 12}$ and ${\bf
w}_{n+\frac 12}$ in Eqs. (\ref{Elfcn}b) and (\ref{Elfcn}c). It is
clear, however, that, if not additionally controlled,  the ideal ring
geometry specified by Eqs. (\ref{Ecnco}) would degrade
because of the approximation errors in Eqs. (\ref{Elfcn}). There are
numerous case-specific ways to handle this problem and some of them
are considered below.

\subsection*{Implementation for Five-Membered Rings}
In this section we consider an implementation of the above algorithm
with examples of specific solutions of the remaining practical
difficulties, such as the construction of a correctly closed loop
conformation and calculation of the projection operator. Our present
implementation is specifically suited for five-membered rings and has
been tested for proline-rich polypeptides and nucleic acids. Cystine
bridges in proteins present a somewhat different case and should better
be treated separately.

Consider once more Fig.\ref{Floop}. It is known since the first
analysis by G{\= o} and Scheraga\cite{Go:70} that the problem of loop
closure in internal coordinates is generally reduced to
six coupled equations. Loop {\em (b)} in Fig.\ref{Floop} shows where this number
comes from. If the loop is broken as shown, each half of the broken
rigid body has six degrees of freedom. In order to close the loop, six
rigid body coordinates of the two parts must be equated. Structure {\em
(c)} in the same figure shows, however, that the complexity of the problem
can sometimes be reduced just by choosing a different underlying tree.
We see that, if all angles are variable, closure in loop {\em (c)}
needs only one distance constraint. The number of
constraining conditions is always equal to the number of degrees of
freedom taken from the underlying tree by the loop closure. All
three loops in Fig. \ref{Floop} are similar, but the underlying tree in
{\em (b)} has five degrees of freedom more than in {\em (c)}. For our algorithm
construction {\em (c)} is certainly preferable.

Let us now turn to five-membered rings. It is known that their
internal motions are well described as pseudorotation with only two
parameters \cite{Altona:72,Gabb:95}. Pseudorotation equations can give
constraints Eqs. (\ref{Ecnco}) in an explicitly inverted form, which
simplifies calculations. However, such small rings are flexible
only if valence angles vary and, therefore, any underlying tree
has rather many degrees of freedom. Thus, the
number of scalar constraints effectively introduced by the
pseudorotation approach is very large even compared with loop {\em (c)} in
Fig.\ref{Floop}. On the other hand, as we just have seen, this
number, in principle, can be reduced to one per ring, and this appears
rather easy.

\begin{figure}
\centerline{\psfig{figure=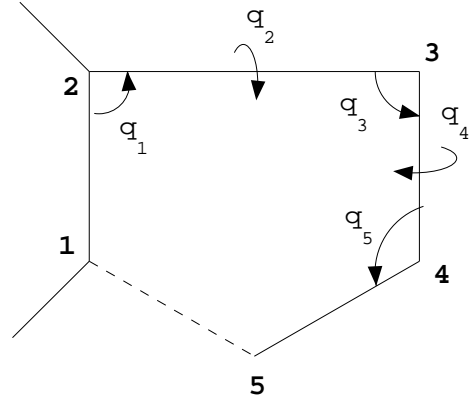,width=7.5cm,angle=0.}}
\caption{\label{F5-mem}
The underlying tree of a five-membered ring. Atoms are
numbered 1,...,5 corresponding to the natural tree ordering. All bond
lengths are fixed. Shown by arrows are five internal coordinates of
the underlying tree that determine the ring conformation.
}\end{figure}

Figure \ref{F5-mem} shows the underlying tree for a five-membered ring.
Atoms are numbered 1,..,5, with this ordering in ribose or deoxyribose
corresponding to
${\rm C'_4,\ C'_3,\ C'_2,\ C'_1,\ O'_4}$ and the broken bond ${\rm
C'_4 ... O'_4}$.
The ring conformation is determined by five valence and dihedral angles
$q_1,...,q_5$ indicated by arrows. Let ${\bf r}_1,...,{\bf
r}_5$ denote atom position vectors and $l_{ij}$ and $e_{ij},\ i,j=1,...,5$
denote the interatomic distances and the corresponding unit vectors.
Directions and positions of rotation axes of ring variables are specified
by the unit vectors ${\bf e}_1,...,{\bf e}_5$ and 
position vectors ${\bf r}_{m_1},...,{\bf
r}_{m_5}$, respectively. The constraint condition is
\begin{equation}\label{E5C=}
C=\left|{\bf r}_5-{\bf r}_1\right|-
l_{15}=0
\end{equation}
We consider firstly calculation of the projection operator ${\bf T}$.
Each ring contributes to the basis matrix ${\bf G}$ a single vector
${\bf g}$ with only five non-zero components obtained by straightforward
differentiation of Eq. (\ref{E5C=}):
\begin{equation}\label{E5g=}
g_i=\frac{\partial C}{\partial q_i}=
{\bf e}_{15}\cdot{\bf e}_i\times\left({\bf r}_5-{\bf r}_{m_i}\right).
\end{equation}
These computations are sufficient to evaluate {\bf T}. In
practice it is used in Eq. (\ref{Elfcn}d) in the factorized form
of Eq. (\ref{ET=}), which
results in several matrix-vector multiplications. Only the term
$\left({\bf G}^*{\bf M}^{-1}{\bf G}\right)^{-1}$ needs to be computed
separately, and the cost of these computations appears minor. The
product $k\times k$ matrix is small and essentially diagonal because
constraints in different rings are only weakly coupled. In addition,
I have found that this term converges faster than the overall iterative
cycle, Eqs. (\ref{Elfcn}b-d), and there is no need to recalculate it
after the first two iterations.

The small deviations of the constrained bond lengths $l_{15}$ caused
by the approximation errors can be either eliminated by exact
analytical ring closure\cite{BKS1:89,Mzc&c:90,Lugovskoy:72}
or reduced to a low and stable level by
feedback corrections.\cite{LINKS:97} The analytical
closure is simple. Let us take variables $q_1,...,q_4$ as independent
and compute the last valence angle $q_5$ so that Eq. (\ref{E5C=}) is
fulfilled. Variables $q_1,...,q_4$ specify positions of atoms 1,...,4
and the orientation of the plane of $q_5$ specified by vectors ${\bf
e}_{34}$ and an in-plane unit vector ${\bf e}_{345}$ orthogonal to
${\bf e}_{34}$. We may write
\begin{equation}\label{E5r45=}
{\bf r}_{45}=l_{45}\left(x{\bf e}_{34}+y{\bf e}_{345}\right)
\end{equation}
where $x$ and $y$ are the two unknown in-plane coordinates of the unit
vector ${\bf e}_{45}$ They are found from the constraint
Eq. (\ref{E5C=}) and the normalization condition, which results in
{\mathletters\label{E5sys}\begin{eqnarray}
x\left({\bf e}_{14}{\bf e}_{34}\right)+y\left({\bf e}_{14}{\bf
e}_{345}\right)&=&\frac{l_{15}^2-l_{45}^2-l_{14}^2}{2l_{14}l_{45}}\\
x^2+y^2&=&1
\end{eqnarray}}
This system is reduced to a square equation and gives a single $x>0$
solution, which solves the task. Derivatives of the dependent angle
$q_5$, which are normally used only for computing energy gradients
during minimization, are
$$\frac{\partial q_5}{\partial q_i}=-g_i/g_5$$

The feedback algorithm is constructed as follows. Suppose, at the nth
step, we have a non-zero value $C_n$ in Eq. (\ref{E5C=}). We require that
\begin{equation}\label{E5Cn+}
C_n+\frac{\partial C}{\partial{\bf q}}{\bf\dot q}_{n+\frac 12}h=0,
\end{equation}
which means that at the next time step the accumulated error must be zeroed
to the first order. It is clear that for each ring only the
component of ${\bf\dot q}_{n+\frac 12}$ orthogonal to the constraint
hypersurface matters
and from Eq. (\ref{E5Cn+}) it is evaluated as
\begin{equation}\label{E5dq=}
{\bf\dot q}_{n+\frac 12}^\bot=-\frac{C_n}{|{\bf g}|^2h}{\bf g}.
\end{equation}
Vectors $\bf g$ are mutually orthogonal
and by combining Eqs. (\ref{E5dq=}) for all rings we obtain a corresponding
component ${\bf\dot q}_{n+\frac 12}^\bot$ for the whole molecule.
To compute the correcting velocity
${\bf\dot q}_{n+\frac 12}^c$ we require that it result from a variation
of reactions and thus
belongs to the subspace with the basis matrix ${\bf M}^{-1}{\bf G}$.
This gives a projection
\begin{equation}\label{E5dqc=}
{\bf\dot q}_{n+\frac 12}^c={\bf\tilde T\dot q}_{n+\frac 12}^\bot=
{\bf M}^{-1}{\bf G}\left({\bf G}^*{\bf M}^{-1}{\bf G}\right)^{-1}
{\bf G}^*{\bf\dot q}_{n+\frac 12}^\bot.
\end{equation}
The last computation tends to reduce the extra mechanical work
introduced by the algorithm.
The correction obtained may be used inside the
iterative cycle of the algorithm (\ref{Elfcn}) or just added to
${\bf\dot q}_{n+\frac 12}$ at the end. Also,
${\bf\dot q}_{n+\frac 12}^\bot$ can be computed for on-step or for
half-step conformations, or both
such vectors may be combined. These alternatives give a series of
slightly different algorithms which are compared in the numerical tests
below. Note, however, that both the analytical closure and the feedback schemes
break the time reversibility of algorithm (\ref{Elfcn}) and should generally
increase the drift of the total energy.

\subsection*{Numerical Examples}
In the numerical tests presented below we address two issues. First,
we test the accuracy of the ring closure provided by the new algorithm
with and without additional corrections. Second, we check its
stability with elevated time steps, notably, the possibility of step
sizes around 10 fsec for proteins and nucleic acids. This specific
value is targeted because it has been found optimal, in a certain
sense, for in-water simulations of proteins\cite{Mzjpc:98}.

We consider two test examples: one for proteins and one for nucleic
acids. The first is a 36 residue fragment of a collagen triple helix
which involves 24 prolines (file 1bff\cite{Nemethy:92} in the protein
database.\cite{PDB:}) Parameters for the amino acids were taken from
the AMBER94 set\cite{AMBER94:}. Except for prolines, all bond lengths and
bond angles were fixed at standard values according to the standard
geometry approximation.\cite{ECEPP2:75} The second test system is a
decamer DNA duplex (TA)$_5$. Nucleotide geometry was taken from FLEX
force field\cite{JUMNA:95} to provide compatibility with JUMNA program
\cite{JUMNA:95} which was used for preparation of the initial duplex
conformations. Except for sugar rings, the geometry of the nucleotides was
fixed, that is the bases were rigid and all other bond lengths
and bond angles were fixed. Calculations have been performed without
explicit solvent by using the AMBER94 force field\cite{AMBER94:} with a
dielectric constant $\epsilon=r$ and phosphate charges in DNA reduced
by 0.5. These conditions emulate the effects of ion condensation and
provide a reasonably accurate approximation in conformational analysis
of nucleic acids.\cite{Flatters:97}

Pyrrolidine and furanose rings are treated similarly by a
single program. The furanose underlying tree has been detailed above. In
the analogous construction for prolines the NC$_\delta$ bond is broken
and replaced by a distance constraint. Our approach generally allows
for arbitrary freezing of internal coordinates and thus can consider
numerous different representations of such rings\cite{BKS0:89}. Here
we consider a model with only bond lengths fixed and all intra-cyclic valence
angles free. These choices leave several fast bond
angle bending modes active, notably, the scissor H-C-H mode with a
frequency of around 1500 cm$^{-1}$. Theory\cite{Mzjcp:97} says
that for a step size of 10 fsec, the maximum frequency should
be 3 times lower. To achieve this, additional moments of inertia of 9
amu$\cdot$\AA$^2$ are applied to C-H bonds as described earlier.
\cite{Mzjcc:97} Following earlier conclusions\cite{Mzjcc:97,Mzjpc:98}
other hydrogen-only rigid bodies, like thymine methyls, have additional
inertia of 4 amu$\cdot$\AA$^2$ added.

\begin{figure}
\centerline{\psfig{figure=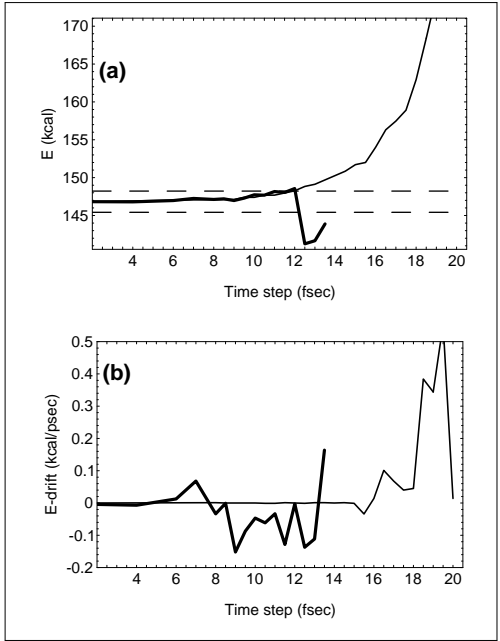,height=8.5cm,angle=0.}}
\caption{\label{Fta10}
Characteristic time step dependencies of the total energy (a)
and of the E-drift (b) for a decamer DNA duplex (TA)$_{10}$. Low
temperature plots are shown by thinner lines. For comparison, in (a)
the low temperature deviations are scaled and shifted to fit the range
of deviations observed with normal temperature. Similarly, in (b) the
low temperature E-drift is multiplied by 100. The dashed horizontal
lines in (a) show the band of acceptable deviation defined in the
text.}\end{figure}

With these modifications, the pseudorotation normal modes with
frequencies around 550 and 640 cm$^{-1}$ for DNA and collagen,
respectively, become the fastest and they already correspond to a
harmonic characteristic time step between 9 and 10 fsec for
leapfrog-equivalent integrators.\cite{Mzjcp:97} However, with
unfavorable collision angles, non-hydrogen ring atoms with all valence
angles free have a considerably smaller effective inertia than similar
atoms in models with fixed bond angles.\cite{Mzjpc:98} In preliminary
tests (not shown here) I observed an anharmonic limitation below
10 fsec at normal temperature which could be overcome by an additional
increase of the ring inertia. Here I show results obtained with the moments
of ring C-C and C-O bonds increased by 15 amu$\cdot$\AA$^2$ and 2
amu$\cdot$\AA$^2$, respectively. This should make the inertias of all ring
bonds similar and approximately equal to that of a water molecule,
with a 50\% increase for C-C bonds. As a side effect, the fastest
pseudorotation frequencies are shifted down below the already lowered
hydrogen scissor modes, and the latter thus remain the fastest in both
test systems.

\begin{figure}
\centerline{\psfig{figure=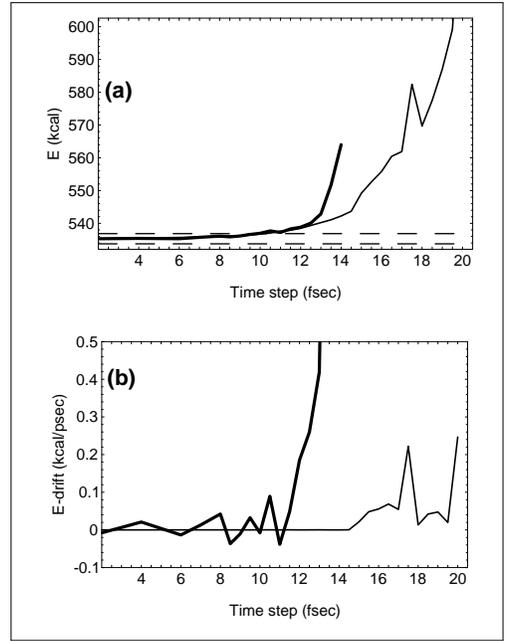,height=8.5cm,angle=0.}}
\caption{\label{Fcoll}
Characteristic time step dependencies of the total energy (a)
and of the E-drift (b) for a dodecamer collagen triple helix.
Notation is same as in Fig. \protect\ref{Fta10}.
}\end{figure}

Stability and step size limits were evaluated with the testing
technique and equilibration protocols proposed and analyzed
in detail elsewhere.\cite{Mzjpc:98,Mzjcp:97} In this method, the test
trajectory is repeatedly calculated always starting from exactly the
same constant-energy hypersurface. In each run, certain system averages
are evaluated and compared with ``ideal'' values, i.e. the same
parameters obtained with a very small time step. The choice of such
parameters has been discussed in detail earlier.\cite{Mzjpc:98} Here
we use only the total energy, $E=\bar U+\bar K$ and its drift
($E$-drift), where $\bar U$ and $\bar K$ are average potential and
kinetic energies computed for integer steps and half-steps,
respectively. As in the previous study\cite{Mzjpc:98} we take a
deviation of $0.2D[U]$, where $D[.]$ denotes operator of variance, as
the upper acceptable level for deviation of $E$. The step size maximum
determined is denoted as $h_{c}$ and called ``characteristic''. The
$E$-drift is exactly zero for ideal harmonic systems,\cite{Mzjcp:97}
and is thus a good indicator of anharmonic effects. Virtually harmonic
conditions are simulated by reducing the temperature down to 0.1K with
the same equilibration protocol as before.\cite{Mzjpc:98} Relevant
harmonic frequencies were evaluated from low temperature spectral
densities of autocorrelation functions of appropriate generalized
velocities. In all production runs the duration of the test trajectory
was 10 psec.

The results of such testing are shown in Figs. \ref{Fta10} and
\ref{Fcoll}. For both model systems the low and normal temperature
plots have characteristic qualitative differences,\cite{Mzjpc:98} but
$h_c$ does not change with temperature indicating that the time step
limitations are harmonic. The $h_c$ values are close to the expected
harmonic estimate. The small difference observed between the two
systems can be attributed to a three times larger number of
hydrogen scissor modes in collagen. We see that the projection step in
algorithm (\ref{Elfcn}) does not deteriorate the high stability of the
original leapfrog algorithm. We conclude also that our model for five-membered
rings allows calculations with $h\approx10$~fsec. The last
conclusion has been checked by computing several nanosecond
trajectories of different DNA oligomers (results not shown).

Plots in Figs. \ref{Fta10} and \ref{Fcoll} are obtained by algorithm
(\ref{Elfcn}) without corrections of constrained bond lengths. Before
considering the effects of such corrections let us look more carefully at
how constraint distances behave in the above conditions. Since
algorithm (\ref{Elfcn}) keeps no information about the initial bond
lengths a diffusive drift from initial values is possible. Note,
however, that the constrained distances are just additional first
integrals of the constrained equations of motion, like momenta or the
total energy. For leapfrog-equivalent integrators
deviations in first integrals caused by approximation errors are normally
oscillatory rather than diffusive, and the drift may thus be small.

\begin{table*}[t]\caption{\label{T1}The quality of the ring closure
obtained with different correcting strategies. Data from 10 psec
trajectories of (TA)$_5$ starting from the same initial state with
all rings closed exactly. Time step 10 fsec. The rms deviations of
closing bond lengths from ideal values are shown, with corresponding
maximal values given in brackets.}
\begin{tabular}[t]{|c|c|c|c|c|}\hline
Algorithm & on-step rings ($\times 10^{-3}$\AA)
& half-step rings ($\times 10^{-3}$\AA)& E-drift ($\times 10^{-2}$kcal/psec)\\
\hline\hline
No correction & 1.50(11.3) & 0.757(6.34) & -9.23\\\hline
Analytical closure & 0(0) & 0(0) & 3.49\\\hline
Feedback on-step & 0.294(1.84) & 1.52(7.81) & 5.71\\\hline
Feedback half-step & 1.53(12.0) & 0.369(2.90) & -39.5\\\hline
Feedback mixed & 0.837(5.72) & 0.852(6.04) & -8.94\\\hline
\end{tabular}
\end{table*}

\begin{figure}
\centerline{\psfig{figure=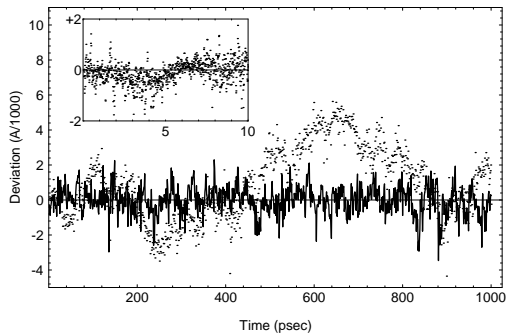,width=7.5cm,angle=0.}}
\caption{\label{Fpro}
Time variation of the constrained bond length in a separate
terminally blocked proline residue during a 1 nsec trajectory. The
dotted plot corresponds to a trajectory computed by algorithm
(\protect\ref{Elfcn}) with no corrections. Its initial part is detailed in the
insertion. The solid line corresponds to a trajectory computed
similarly, but with the analytical ring closure applied once in 10
psec.}\end{figure}

Figure \ref{Fpro} shows the time variation of a CN$_\delta$ bond in proline
during a 1 nsec trajectory. This trajectory has been computed
for a single terminally blocked residue at normal temperature with the
same conditions as above. It is
seen that the fluctuations have many time scales, but even in the
slowest one they are not evidently diffusive. The insertion plot shows
that the high frequency amplitude of the fluctuations levels starting
from the very first time step. It scales as $O(h^2)$ in agreement
with the general properties of leapfrog-like algorithms\cite{Mzjcp:97}
(not shown). The high-frequency amplitude is evidently larger than
diffusive deviations accumulated for tens of picoseconds, and, therefore,
infrequent periodical corrections should be sufficient to keep
deviations within this range. Such a possibility is illustrated by
the solid line in the same figure where the analytical ring closure
was applied once every 10 psec. One may note that this gives a
reasonable and certainly the safest correction strategy.

Table \ref{T1} compares several correcting schemes discussed in the
previous section. We noted above that numerous feedback algorithms are
possible due to variation of two conditions. First, correction of
generalized velocities can be added within the iterative cycle or
after it. The last option, however, always yields a considerably
higher E-drift and such algorithms are not included in Table \ref{T1}.
Second, there are two possible directions of the correcting vector
computed for on-step rings and half-step rings, respectively. Only
one of the two directions, or an average of the two vectors, can be
used, or else they can alternate between time steps. In the last case,
however, a dramatic loss of the overall stability of the algorithm is
induced. Thus, among many variants only three feedback strategies
included in Table \ref{T1} give acceptable results. The data have been
obtained for 10 psec trajectories of the DNA decamer with a 10 fsec step
size. Trajectories started from the same initial state with sugar
rings closed exactly.

We noted above that any correcting algorithm is likely to increase the
E-drift. It appears, however, that the increase is usually below the
noise level for the step sizes of interest, which is clearly seen in Table
\ref{T1} and Fig. \ref{Fta10}. Except for half-step feedback
corrections the E-drift is within the range of fluctuations in Fig.
\ref{Fta10}(b). A certain increase in E-drift is observed,
however, with much smaller time steps as well as with $h>h_c$. Very
good results are obtained with analytical ring closure, which seems to
be the best choice for furanose and proline rings. It should be noted that,
in this case, with any step size, iterations in Eqs. (\ref{Elfcn})
converge only up to a relative accuracy of 10$^{-5}$ and then start
looping. This, however, does not seem to affect either the accuracy
in terms of energy conservation, or the long time stability of the
algorithm, which also has been checked for nanosecond trajectories.

Although feedback algorithms appear unnecessary for five-membered
rings they still present significant interest especially for S-S bridges
were both the analytical closure and the periodical
corrections are not easy. Table \ref{T1} shows that, as expected, the three
feedback algorithms improve the accuracy in the targeted rings.
Improvements are not spectacular, but it is important to note that
the corrected levels of deviations are stable in time. In this
respect the last algorithm in Table \ref{T1}, which manages to correct both
on-step and half-step distances, is the most promising.

\section{Conclusions}
This study is, to my knowledge, the first successful attempt to develop a
practical ICMD approach to large molecules with internal flexible
rings. It is shown here that the strategy referred to above as
``inconsistent'', that is imposing explicit constraints upon a system
already constrained implicitly, results in algorithms as fast and as
stable as those for ICMD simulations of polymers with the tree topology.
For the important case of five-membered rings in nucleic acids and proteins,
calculations with time steps around 10 fsec are shown to be possible.

\acknowledgements
I wish to thank R. Lavery for useful discussions and critical comments to the
manuscript.

\end{document}